\documentclass[pra,twocolumn,superscriptaddress,notitlepage]{revtex4-1}
\usepackage{graphicx,amsmath,amsfonts,amssymb,upgreek,txfonts,color}
\usepackage[colorlinks,linkcolor=blue,citecolor=blue,urlcolor=blue,breaklinks=true]{hyperref}

\begin{document}

\title{Microcavities with suspended subwavelength structured mirrors}

\author{Andreas Naesby}
\author{ Aur\'{e}lien Dantan}

\email[Corresponding author: ]{dantan@phys.au.dk}
\affiliation{Department of Physics and Astronomy, University of Aarhus, DK-8000 Aarhus C, Denmark}

\date{\today}

\begin{abstract}
We investigate the optical properties of microcavities with suspended subwavelength structured mirrors, such as high-contrast gratings or two-dimensional photonic crystals slabs, and focus in particular on the regime in which the microcavity free-spectral range is larger than the width of a Fano resonance of the highly reflecting structured mirror. In this unusual regime, the transmission spectrum of the microcavity essentially consists in a single mode, whose linewidth can be significantly narrower than both the Fano resonance linewidth and the linewidth of an equally short cavity without structured mirror. This generic interference effect---occuring in any Fabry-Perot resonator with a strongly wavelength-dependent mirror---can be exploited for
realizing small modevolume and high quality factor microcavities and, if high mechanical quality suspended structured thin films are used, for optomechanics and optical sensing applications. 
\end{abstract}

\maketitle

\section{Introduction}
Interference effects in structured thin films are widely exploited in photonics to tailor the properties of integrated optical elements. Of particular interest are thin films patterned with subwavelength structures, such as high-contrast gratings~\cite{Chang-Hasnain2012} or photonic crystals~\cite{Zhou2014}. There, interference between modes propagating through the film and transverse guided modes result in the appearance of Fano resonances~\cite{Miroshnichenko2010,Limonov2017}, which can bring about remarkable optical properties, such as broadband high-reflectivity or transmissivity, or the appearance of high quality factor resonances. Such features can be exploited for realizing a wide range of integrated photonics components, e.g., optical filters~\cite{Shuai2013}, couplers~\cite{Zhou2009}, reflectors~\cite{Brueckner2010}, lasers~\cite{Huang2007,Boutami2007,Wagner2016}, detectors~\cite{Chen2010}, sensors~\cite{Zhou2008}, etc.

While such subwavelength structured films inherently display Fabry-Perot--type interferences~\cite{Sauvan2005,Karagodsky2011}, they are typically integrated with standard optical elements, e.g., in linear Fabry-Perot resonator configurations in order to enhance their spectral selectivity, detection sensivity, or the strength of the light-matter interaction~\cite{Chang-Hasnain2012,Zhou2014,Miroshnichenko2010,Limonov2017}.

In this work, we investigate microcavities with suspended subwavelength structured mirrors and focus in particular on the regime in which the microcavity free-spectral range is larger than the width of a Fano resonance of the highly reflecting structured mirror. In this unusual regime, the transmission spectrum of the microcavity essentially consists in a single mode, whose linewidth can be significantly narrower than both the Fano resonance linewidth and the linewidth of an equally short cavity without structured mirror. We show for instance that few-micron long resonators with a Fano mirror with a moderately high-$Q$ resonance of a thousand and having finesse of a few hundreds at $\mu$m wavelengths can display GHz-wide transmission features. This generic interference effect---occuring in any Fabry-Perot resonator with a strongly wavelength-dependent mirror---could thus be exploited to realize small modevolume and high quality factor microcavities. This is highly relevant for a wide range of spectroscopy and light-matter investigations in which it is desirable to increase the strength of the light-matter interaction with  emitters by reducing the cavity modevolume without losing on its spectral selectivity~\cite{Bitarafan2017}. They would also be beneficial for laser applications involving distributed Bragg gratings and ultracompact vertical or composite cavities~\cite{Huang2007,Boutami2007,Wagner2016}.

Furthermore, these remarkably narrow linewidths are shown to be reasonably robust with respect to wavefront curvature, imperfect parallelism and finite size effects, so that such microcavities could be realized in practice in a simple plane-parallel geometry, without resorting to optical elements with short focusing abilities~\cite{Fattal2010,Klemm2013,Arbabi2015,Guo2017}. There, the combination of ultrashort cavities with low-mass and high-mechanical quality \textit{suspended} structured thin films~\cite{Kemiktarak2012,Bui2012,Norte2016,Reinhardt2016,Bernard2016,Chen2017} would be particularly interesting for optomechanics~\cite{Thompson2008,Kemiktarak2012NJP,Xuereb2012,Reinhardt2016,Piergentili2018,Gartner2018} and optical switching applications~\cite{Yang2013,Hui2013}, as it provides a means of increasing radiation pressure forces without reducing the cavity linewidth. They would also be attractive for optical sensing applications in which reduction in the resonator length and high interferometric sensitivity are desirable, e.g., accelerometry and gravitometry~\cite{Krause2012,Cervantes2014,Guo2017,Armata2017,Qvarfort2017} or pressure and ultrasound sensing~\cite{Naesby2017,Leinders2015,Basiri2018}. Finally, realizing such microcavities by integrating the structured mirror in a fiber-optic Fabry-Perot interferometer could be an interesting alternative to fiber sensors with in-fiber embedded Bragg gratings~\cite{Islam2014}.

\begin{figure}[htbp]
\centering
\includegraphics[width=0.7\linewidth]{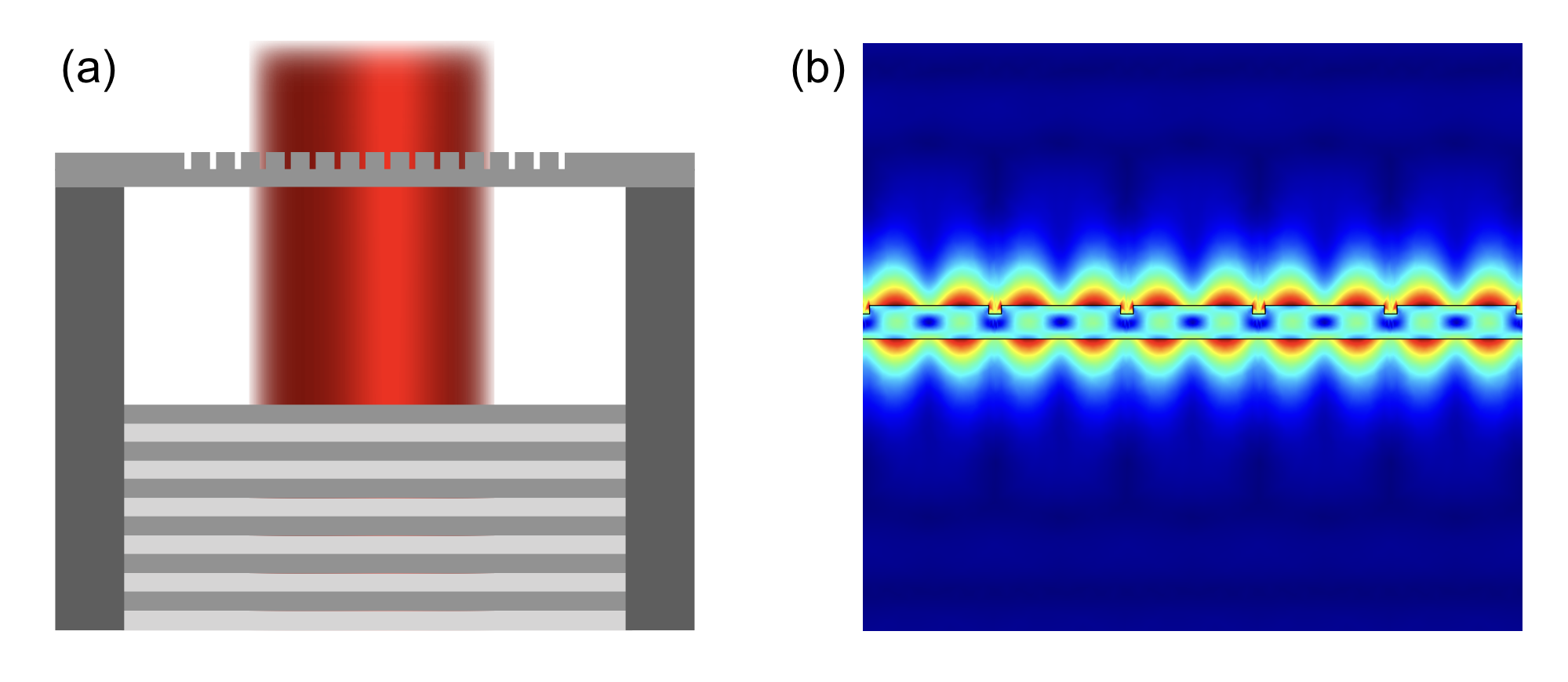}
\caption{(a) Schematic of a microcavity consisting of a suspended Fano mirror (HCG) and a highly reflecting Bragg mirror. (b) Numerically calculated electric field intensity at a Fano resonance of the microcavity simulated in Sec.~\ref{sec:1D}. The black lines indicate the contours of the HCG.}
\label{fig1}
\end{figure}

\section{Idealized 1D model}\label{sec:1D}
We start by considering an idealized one-dimensional scattering model, in which the optical resonator consists of two parallel, absorption-free mirrors (fig.~\ref{fig1}): a highly reflecting mirror with amplitude transmission and reflection coefficients $r$ and $t$, and a ``Fano" mirror, whose reflection and transmission coefficients result of the interference between the direct transmission through the slab and a guided transverse mode~\cite{Fan2002,Fan2003}
\begin{equation}
r_g(\omega)=i\frac{t_d\gamma +r_d(\omega-\omega_0)}{\gamma+i(\omega-\omega_0)},\hspace{0.2cm}t_g(\omega)=\frac{r_d\gamma-t_d(\omega-\omega_0)}{\gamma+i(\omega-\omega_0)}\,,
\label{eq:rgtg}
\end{equation}
where $t_d$ and $r_d$ are the normal incidence transmission and reflection coefficients, and $\gamma$ the width of the transverse guided resonance mode with frequency $\omega_0$. This coupled-mode model is known to accurately reproduce the Fano resonances typically observed with subwavelength high-contrast grating (HCG) or photonic crystal structures~\cite{Miroshnichenko2010,Zhou2014}, and can thus be used as a simple basis to discuss the physics of Fano mirror resonators. For convenience, we start by modelling the mirrors as infinitely thin 1D scatterers, characterized by their polarizabilities $\zeta=-ir/t$ and $\zeta_g(\omega)=-ir_g(\omega)/t_g(\omega)$, respectively. We assume that the highly-reflecting mirror polarizability $\zeta$ does not significantly vary over the frequency range of interest, while the frequency dependence of $\zeta_g(\omega)$ is prescribed by eq.~(\ref{eq:rgtg}).
Denoting by $l$ the cavity length, the overall transmission of the optical resonator is then given by
\begin{equation}
\mathcal{T}=\left|\frac{tt_g(\omega)}{1-rr_g(\omega)e^{i2\omega l/c}}\right|^2\,.
\end{equation}
To simplify the discussion we assume that the cavity length $l$  is chosen such that the guided mode resonance frequency $\omega_0$ coincides with one of the bare optical cavity resonance frequencies, satisfying $\omega_0=(c/2l)(2\pi p+\arctan(1/\zeta))$, with $p$ an integer and $c$ the speed of light in vacuum. We assume $r_g$ and $t_g$ real for simplicity, and take $\zeta(\omega_0)\sim \zeta$ in order to mimic a symmetric cavity.

\begin{figure}[htbp]
\centering
\includegraphics[width=0.6\linewidth]{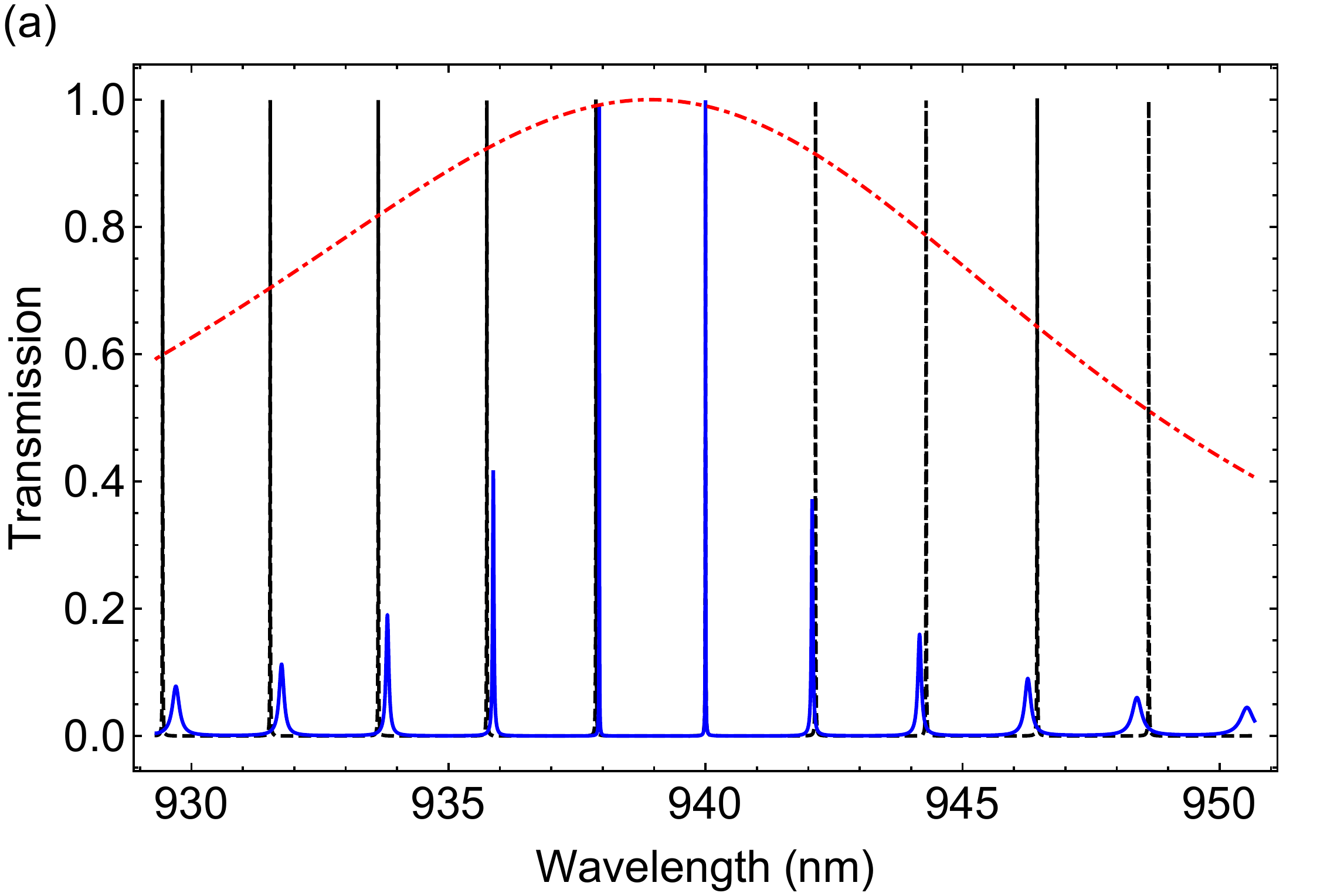}\\
\includegraphics[width=0.6\linewidth]{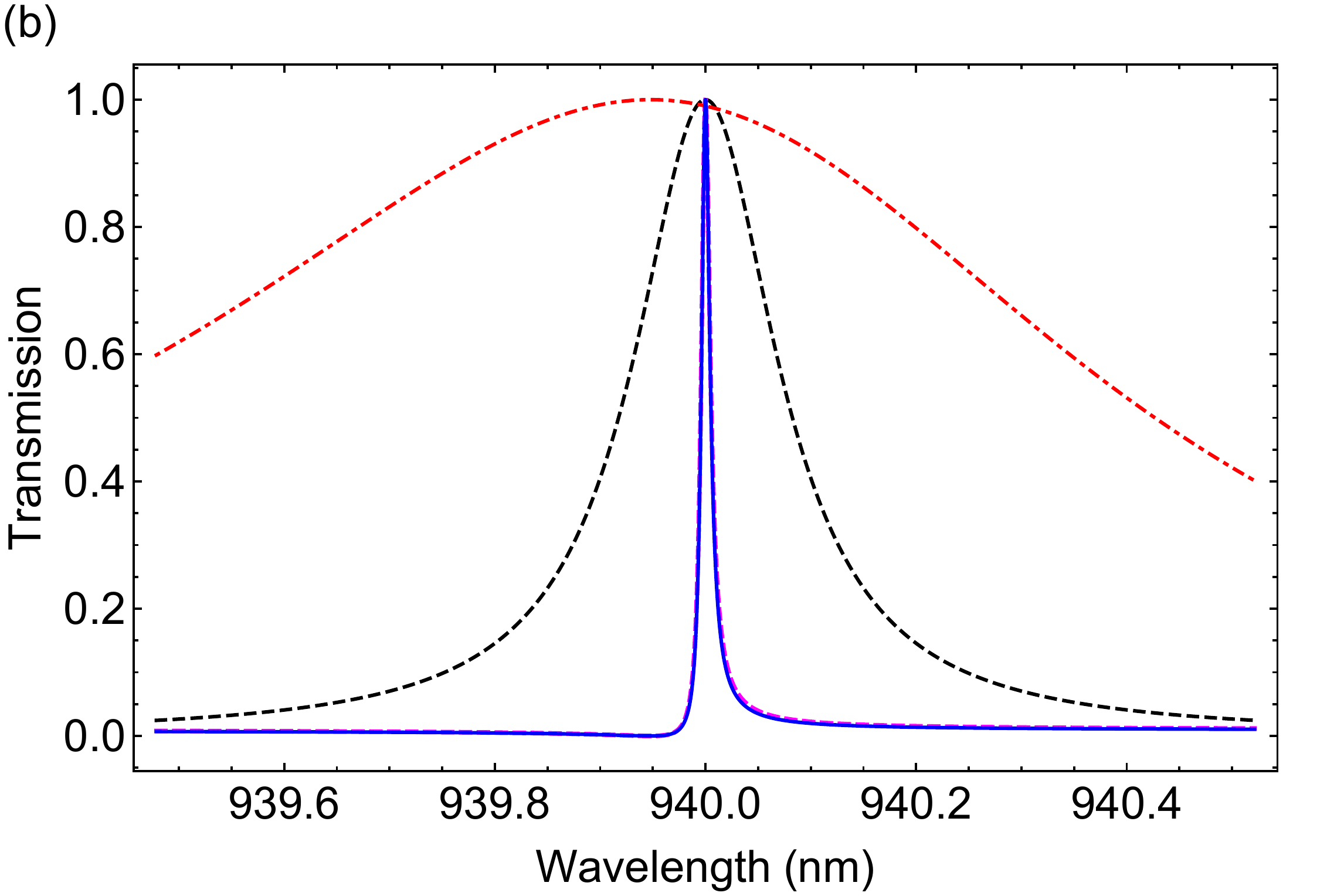}\\
\includegraphics[width=0.6\linewidth]{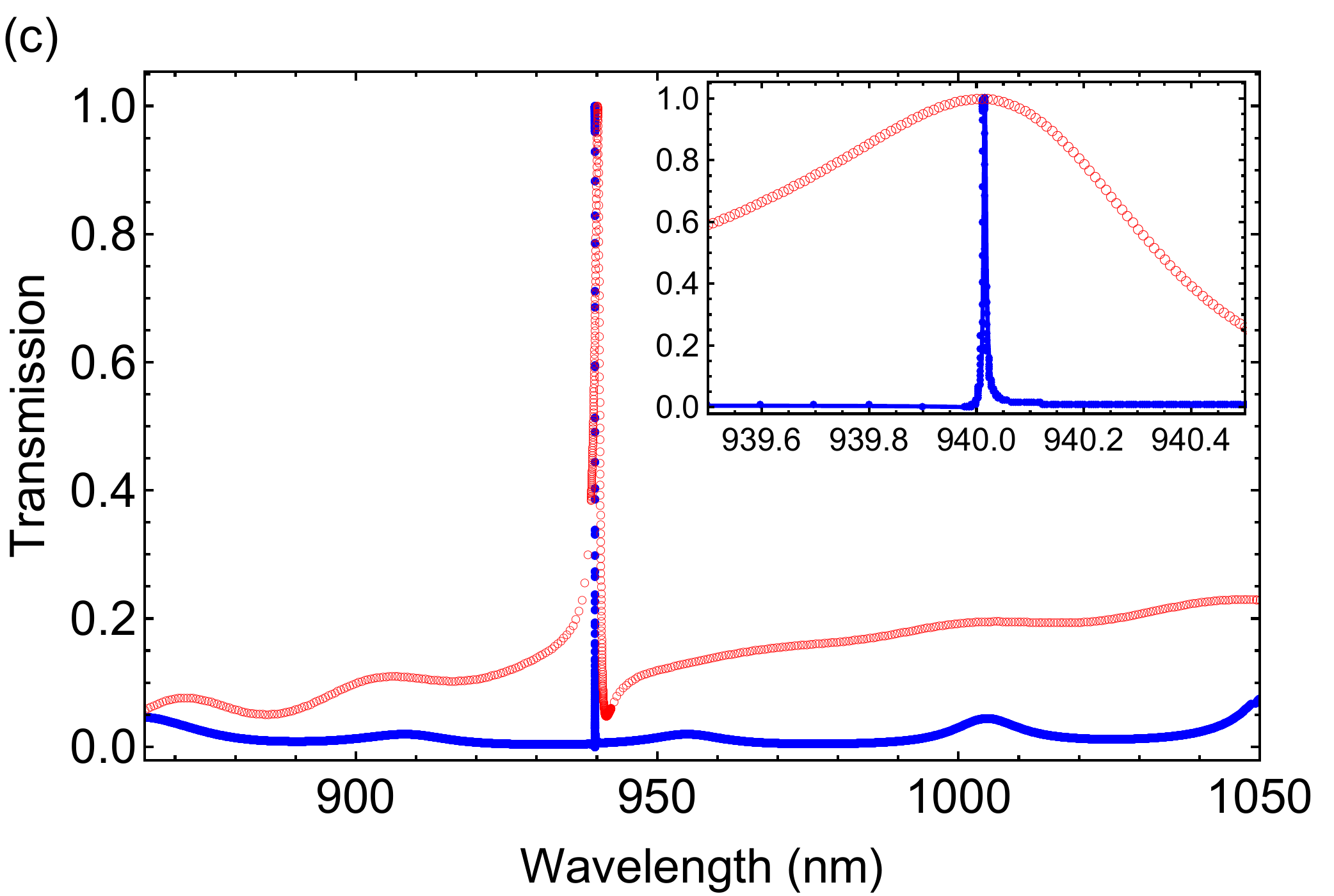}
\caption{Transmission spectrum of (a) 206.8~$\mu$m-long resonator with low-$Q$ Fano mirror resonance of 21.3 ($\gamma=5\gamma_{\textrm{FSR}}$)  and (b) 8.5~$\mu$m-long resonator with high-$Q$ Fano mirror resonance of $900$ ($\gamma=\gamma_{\textrm{FSR}}/100$). Parameters: $|r|^2=|t_d|^2=0.99$, $\lambda_0=2\pi c/\omega_0=940$ nm. The plain blue curves show the results of the transfer matrix calculations, while the dashed red and dashed-dotted black curves show the Fano mirror reflectivity $|r_g|^2$ and the bare cavity spectrum, respectively. The dashed magenta curve in (b) shows the result of eq.~(\ref{eq:Tapprox}). (c) Full-numerical simulation of the transmission spectrum of a realistic Fano cavity (blue dots) consisting of a HCG and a Bragg mirror with parameters similar to (b). The red circles show the HCG reflectivity and the inset a zoom around the 6 pm (2 GHz)-wide resonant transmission line.}
\label{fig2}
\end{figure}

Depending on the ratio of the free spectral range $\gamma_{\textrm{FSR}}=c/(2l)$ and the width of the subwavelength structured mirror resonance $\gamma$, two regimes can be considered. Typically, ``long" cavities and mirrors with relatively broad Fano resonances ($\gamma_{\textrm{FSR}}\ll\gamma$) exhibit, within the bandwidth of the Fano resonance, Lorentzian cavity resonances with a linewidth given by the ``bare" cavity linewidth $\kappa=\gamma_{\textrm{FSR}}/\pi\zeta^2$ (for $\zeta\gg 1$). By the ``bare" cavity we refer here to a cavity having the same length, but for which both mirrors have a frequency-independent polarizability $\zeta$. This regime is illustrated in Fig.~\ref{fig2}a for a $\sim 200$ $\mu$m-long cavity consisting of a highly reflecting mirror with $|r|^2=0.99$ and a Fano mirror with an optical resonance around 940 nm having a $Q$-factor of 20, resulting in a ratio $\gamma=5\gamma_{\textrm{FSR}}$. The spectrum indeed exhibits cavity modes with linewidth $\kappa$ around the Fano mirror resonance, while modes with larger linewidth and lower peak transmission are observed away from $\omega_0$ due to the decrease in reflectivity of the Fano mirror.

However, for a short enough cavity, such that $\gamma_{\textrm{FSR}}\gg\gamma$, the cavity spectrum consists in one mode only, having a resonance frequency close to $\omega_0$. Remarkably, the linewidth of this mode can be much narrower than both the bare cavity linewidth $\kappa$ and the Fano mirror resonance width $\gamma$. Assuming $\zeta_g(\omega_0)=\zeta\gg 1$, the transmission around resonance can be shown to be approximately given by
\begin{equation}
\mathcal{T}(\delta)\simeq\frac{1}{1+F\left(\frac{\delta}{1-\zeta\delta}\right)^2}\,,
\label{eq:Tapprox}
\end{equation}
where $\delta=(\omega-\omega_0)/\gamma$ and $F\simeq\zeta^4$ is the coefficient of finesse of the bare cavity. The resulting Fano resonance profile displays a linewidth $\gamma/\sqrt{F}$, which is narrower than that of the Fano mirror by a factor given by the coefficient of finesse of the cavity. This effective narrowing in presence of the highly-reflecting cavity mirror can be understood by realizing that, even though the Fabry-Perot resonator only possesses one resonant mode, constructive interference occurs only for photons whose frequency detuning after $\sim \sqrt{F}$ roundtrips from the Fano resonance frequency is less than $\gamma$. This phase sensitivity enhancement can also be seen as a general feature of coupled resonant cavity systems~\cite{Golub2006,Smith2003}. Figure~\ref{fig2}b shows the transmission of an 8.5 $\mu$m-long cavity with a Fano mirror with $Q\sim 10^3$, such that $\gamma=\gamma_{\textrm{FSR}}/100$. The resulting transmission linewidth is $\sim 10$ pm, substantially narrower than the bare cavity linewidth of 0.26 nm and the Fano mirror width of $1.04$ nm, and its profile is seen to be accurately reproduced by eq.~(\ref{eq:Tapprox}).

To assess the results obtained with the infinitely thin scatterer model we numerically simulate the transmission spectrum of a cavity similar to that of Fig.~\ref{fig2}b, consisting in a HCG and a multilayer Bragg mirror as in Fig.~\ref{fig1}. We base ourselves on the HCG structures patterned on suspended sillicon nitride films~\cite{Kemiktarak2012,Kemiktarak2012NJP,Nair2018}, but note that similar results could readily be obtained with photonic crystal structures~\cite{Bui2012,Norte2016,Reinhardt2016,Bernard2016,Guo2017,Chen2017,Moura2018,Gartner2018}. We consider a 200 nm-thick silicon nitride slab (refractive index $n=2.0$), in which a high contrast grating with a period of 779 nm and 705 nm-wide, 50 nm-deep rectangular grating fingers is etched, in combination with a Bragg mirror consisting in 18 alternate layers of SiO$_2$ ($n = 1.455$) and Ta$_2$O$_5$ ($n = 2.041$) with $\pi/2$ thickness, and simulate the field propagation (TM polarisation) using the finite element modeling software Comsol using periodic Floquet boundary conditions. In this way, we realistically simulate a Fano mirror and a highly-reflecting mirror with parameters close to those of Fig.~\ref{fig2}b. For a cavity length of $8.59$ $\mu$m, a $\sim 6$ pm-wide ($\sim 2$ GHz) transmission line is obtained (Fig.~\ref{fig2}c), thus supporting the previous analysis.  Let us also point out that the numbers chosen in this example for the reflectivity level (99\%) and optical quality factor ($10^3$) of the Fano mirrors are quite realistic experimentally~\cite{Kemiktarak2012,Reinhardt2016}. Substantially narrower linewidths could in principle be obtained, were one to use even more reflecting mirrors~\cite{Kemiktarak2012NJP,Chen2017} and/or ultrahigh-$Q$ Fano resonances~\cite{Zhou2008,Shuai2013}.

\section{Transverse effects}

\begin{figure}[htbp]
\centering
\includegraphics[width=0.49\linewidth]{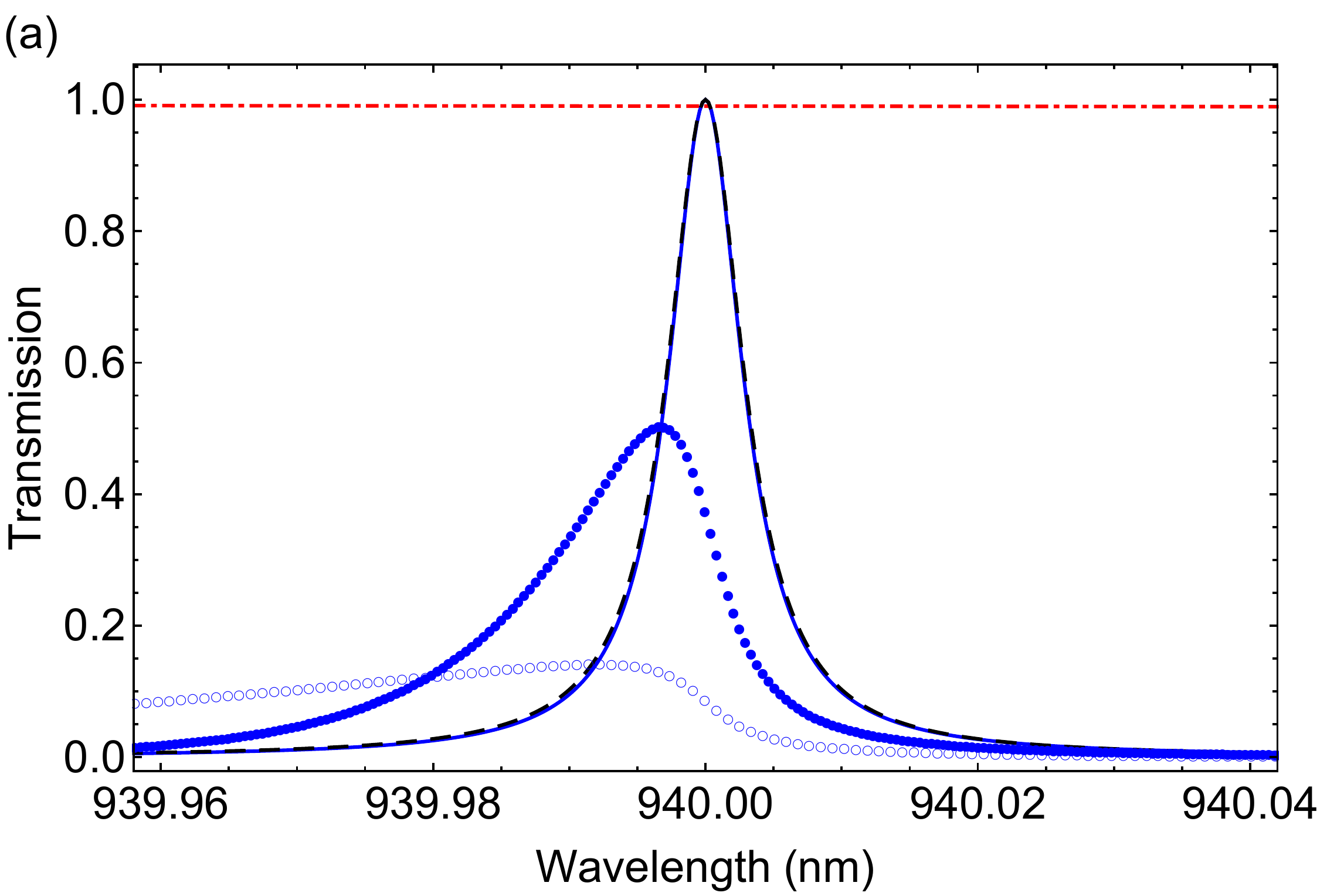}
\includegraphics[width=0.49\linewidth]{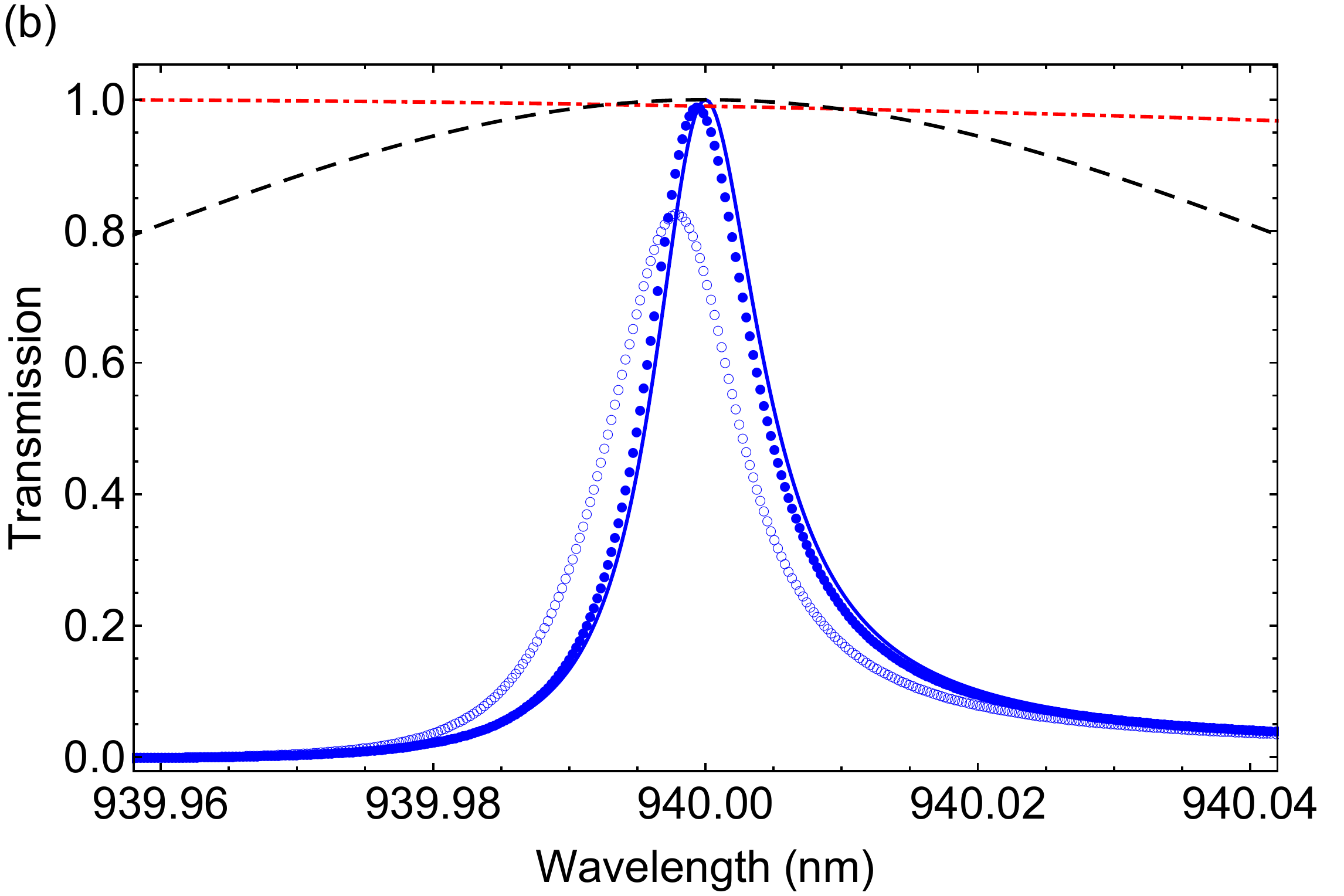}
\caption{(a) and (b): Transmission spectra of the cavities of Fig.~\ref{fig2}a and~\ref{fig2}b, respectively, taking into account the Gaussian wavefunction of the incoming light field. Plain (blue): plane wave case, full dots:  $w_0=50$ $\mu$m, empty circles: $w_0=20$ $\mu$m. For reference, dashed (black): bare cavity spectrum, dash-dotted (red): Fano mirror reflectivity. }
\label{fig3}
\end{figure}

We now address effects going beyond the 1D scenario and start by investigating the effect of the incoming field transverse wavefront on the microcavity linewidth. For perfectly parallel plane mirrors, taking the wavefront curvature into account sets a fundamental limit to the achievable finesse and linewidth of the resonator. Such wavefront effects are particularly relevant for Fano mirrors whose structured area is relatively small, thus constraining the incoming beam size to avoid diffraction losses and, thereby, increasing the beam divergence inside the microcavity. For the sake of concreteness we assume an incoming Gaussian TEM$_{00}$ mode having its waist $w_0$ at the Fano mirror. We take $w_0$ to be much smaller than the structured area in order to neglect trivial diffraction effects. Under these assumptions the outgoing field amplitude after the highly reflecting mirror is given by an infinite sum of reflected components
\begin{equation}
E(r,l)=\sum_n t t_g (r r_g)^ne^{ikz_n}\exp\left(-\frac{r^2}{w(z_n)^2}\right)\exp\left[ik\frac{r^2}{2R(z_n)}-i\psi(z_n)\right]\frac{\sqrt{2/\pi}}{w(z_n)}\,,
\end{equation}
where $k=2\pi/\lambda$, $z_n=(1+2n)l$, $R(z)=z[1+(z_R/z)^2]$, $z_R=\pi w_0^2/\lambda$, $w(z)=w_0\sqrt{1+(z/z_R)^2}$ and $\psi(z)=\arctan(z/z_R)$ are the standard Gaussian beam parameters. The relative transverse dephasing and reduction in reflectivity, which increase for each roundtrip, will obviously lead to a broadening of the spectrum and reduced cavity transmission, as compared to those observed in the idealized 1D plane-wave model.  The normalized cavity transmission $\mathcal{T}=\int_0^{\infty}|E(r,l)|^22\pi rdr$ is plotted on Fig.~\ref{fig3} for the same cavities as in Fig.~\ref{fig2} and for two different waist sizes of 20 and 50 $\mu$m. Figure \ref{fig3}a shows that the transmission of the long cavity is substantially broadened and reduced, even for the larger waist of 50 $\mu$m which corresponds to a Rayleigh range of 8 mm. In contrast, the linewidth of the short cavity is only slightly affected by wavefront curvature effects (Fig.~\ref{fig3}b). This highlights another practical benefit of ultrashort Fano microcavities, for which ultranarrow transmission lines can be obtained even without resorting to Fano mirrors with focusing abilities~\cite{Fattal2010,Klemm2013,Arbabi2015,Guo2017}.

\begin{figure}[htbp]
\centering
\includegraphics[width=.6\linewidth]{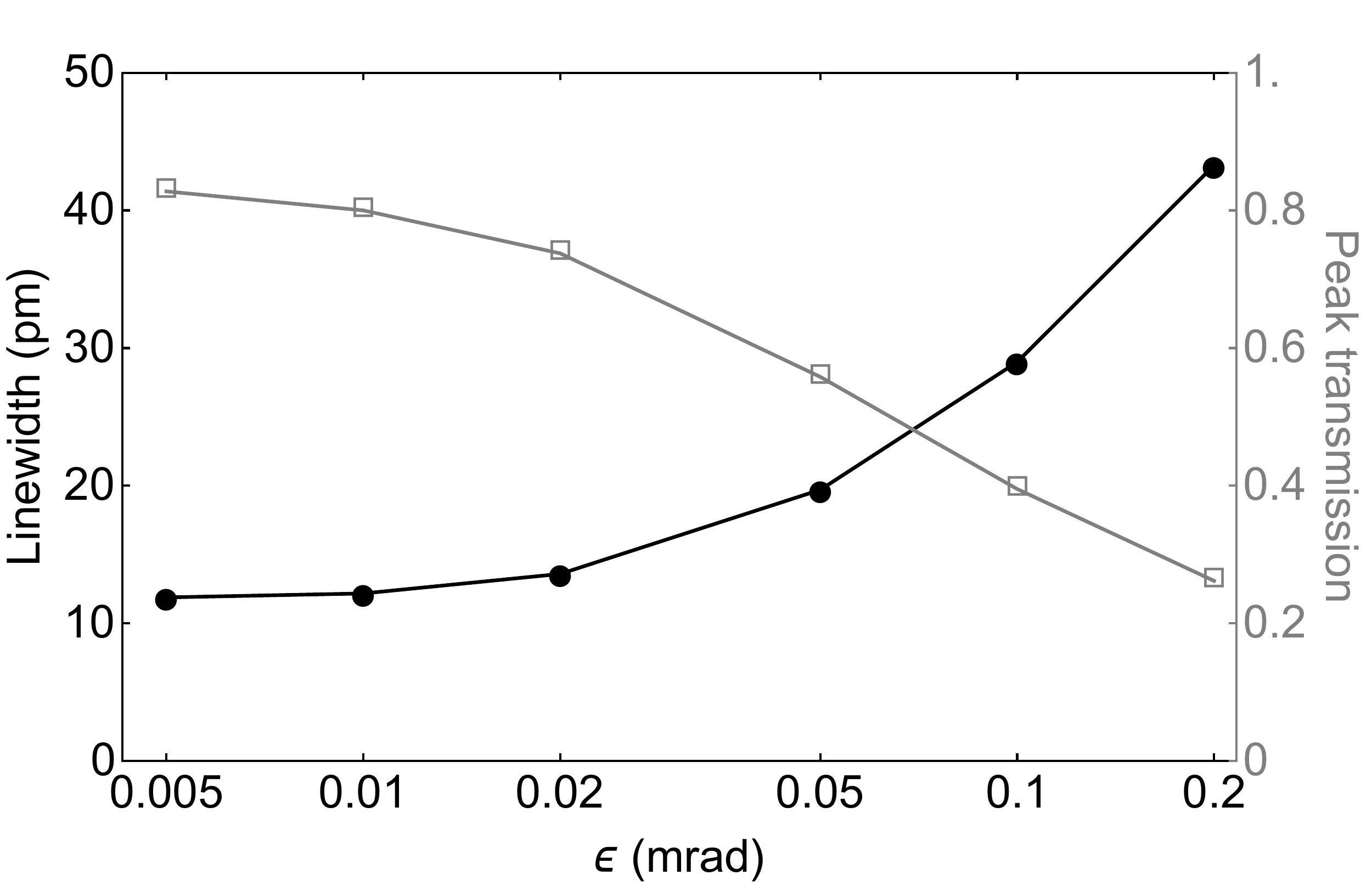}
\caption{Resonance linewidth (black dots) and resonant transmission level (gray squares) of an $8.5$ $\mu$m-long cavity with a $Q=900$ Fano mirror, for various tilt angles $\epsilon$. $w_0=20$ $\mu$m and $|r|^2=|t_d|^2=0.99$. The lines are guides for the eyes.}
\label{fig4}
\end{figure}

Another transverse effect worth investigating is the sensitivity of the microcavity linewidth with respect to imperfect parallelism between the mirrors. We still assume that the previous Gaussian mode impinges at normal incidence ($z$-direction) on the Fano mirror, but now the highly reflecting mirror makes an angle $\epsilon$ with the Fano mirror plane in the, say, $x$-direction. Following the approach of Ref.~\cite{Lee2002}, geometrical considerations for the reflected field amplitudes lead to an outgoing field amplitude after the highly reflecting mirror given by
\begin{equation}
E(x,y,l)=\sum_nt t_g (r r_g)^n E_n(x-x_n,y,l+z_n)\,,
\end{equation}
where $E_n(x,y,z)=E_{\textrm{in}}(x\cos(2n\epsilon),y,z+x\sin(2n\epsilon))\sqrt{\cos(2n\epsilon)}$ is the field amplitude having experienced a wavefront tilt by $2n\epsilon$, $x_n=l/\tan(\epsilon)(1/\cos(2n\epsilon)-1)$ is the transverse walk-off of the $n$-th outgoing beam, $z_n=l\tan(2n\epsilon)/\tan(\epsilon)$ is the distance travelled by the $n$-th outgoing beam with reference to the direct transmission beam ($n=0$). $E_{\textrm{in}}$ is the incoming Gaussian modefunction at the Fano mirror. Figure~\ref{fig4} shows the resonance linewidth and resonant transmission levels of the previous $8.5$ $\mu$m-long cavity with a $Q=900$ Fano mirror resonance, for an incoming beam waist of 20 $\mu$m and for tilt angles which are achievable by, e.g.,  assembly of commercial silicon nitride membranes~\cite{Nair2017}. The linewidth broadening and resonant transmission reduction are consistent with the observation that, for small tilt angles, tilt-induced beam walkoff corrections scale as $F(\epsilon/\theta)^2$, where $\theta=\lambda/\pi w_0$ is the Gaussian beam divergence~\cite{Nair2017}. For a given degree of parallelism and cavity finesse, there generally exists an optimal waist size which minimizes the linewidth, as too large a beam increases the dephasing due to the tilt-induced beam walkoff, but too small a beam results in stronger wavefront effects. Let us finally note that these effects, in particular for high-finesse cavities, would be mitigated by the use of Fano mirrors with focusing abilities~\cite{Fattal2010,Klemm2013,Arbabi2015,Guo2017} or a well-controlled monolithic nanofabrication process~\cite{Gartner2018}.

\begin{figure}[htbp]
\centering
\includegraphics[width=.6\linewidth]{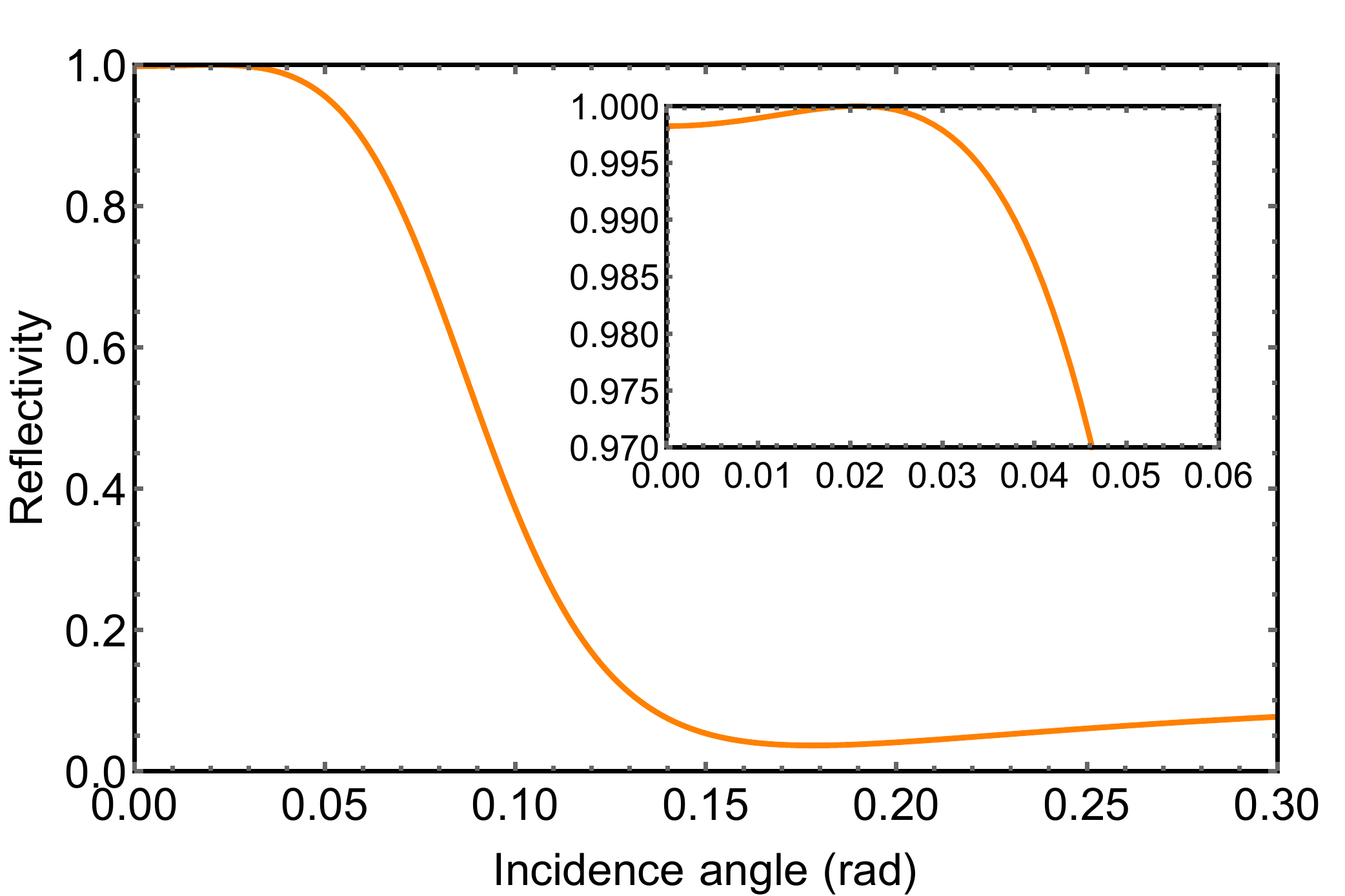}
\caption{Reflectivity at the resonant wavelength of 939.8 nm of the HCG simulated in Fig.~\ref{fig2}c, as a function of the incidence angle. The inset shows a zoom on the small angle/high reflectivity region.}
\label{fig5}
\end{figure}

Last, as abovementioned, the finite lateral size of the structured area may limit in practice the beam waist size. However, as the waist  is reduced, the divergence of the beam increases and coupling to other modes in the structure may induce extra loss and broaden the Fano resonance of the structured mirror itself~\cite{Bui2012,Kemiktarak2012NJP,Norte2016,Bernard2016}. To estimate the minimum waist size required for such couplings not to play a role in a realistic structure we follow Ref.~\cite{Kemiktarak2012NJP} and make use of Rigorous Coupled Wave Analysis simulations~\cite{MIST} to calculate the variations of the  reflectivity at the Fano resonance wavelength of 939.8 nm of the HCG simulated in Fig.~\ref{fig2}c as a function of the incidence angle of TM-polarized plane waves illuminating the infinite grating structure. These variations are shown in Fig.~\ref{fig5}; it can be seen that the reflectivity reaches a 99\% level for an angle $\theta\simeq 0.04$ radians. Equating this angle with the far-field diffraction angle of a Gaussian beam $\theta\simeq \lambda/(\pi w_0)$ yields a minimum waist value of $w_0\simeq 7.5$ $\mu$m. For waists larger than this value, as considered above, such finite-size effects are thus expected to play a negligible role.

\section{Conclusion}
To conclude, we investigated the generic transmission properties of a linear Fabry-Perot resonator incorporating a Fano mirror with a strongly wavelength-dependent reflectivity, such as can be realized with high-constrast gratings or two-dimensional photonic crystals. We showed in particular that enhanced spectral resolution can be achieved with ultrashort microcavities using suspended structured membranes in a simple parallel-plane geometry and for realistic parameters. Cavities with vibrating Fano mirrors would be particularly interesting as integrated devices for photonics and sensing, and, if further combined with electrical actuation~\cite{Nair2018,Wu2018}, for nano-electro-optomechanics~\cite{Midolo2018}.

\section*{Funding}
The Velux Foundations.


\begin{thebibliography}{99}
\bibitem{Chang-Hasnain2012} C. J. Chang-Hasnain and W. Yang, {\it High-contrast gratings for integrated optoelectronics}, Adv. Opt. Photon., {\bf 4}, 379 (2012).
\bibitem{Zhou2014} W. Zhou, D. Zhao, Y.-C. Shuai, H. Yang, S. Chuwongin, A.Chadha, J.-H. Seo, K. X. Wang, V. Liu, Z. Ma, and S. Fan, {\it Progress in 2D photonic crystal Fano resonance photonics}, Progress in Quantum Electronics {\bf 38}, 1 (2014).
\bibitem{Miroshnichenko2010} A. E. Miroshnichenko, S. Flach, and Y. S. Kivshar, {\it Fano resonances in nanoscale structures}, Rev. Mod. Phys. {\bf 82}, 2257 (2010).
\bibitem{Limonov2017} M. F. Limonov, M. V. Rybin, A. N. Poddubny, and Y. S. Kivshar, {\it Fano resonances in photonics}, Nature Photon. {\bf 11}, 543 82017).
\bibitem{Shuai2013} Y. Shuai, D. Zhao, Z. Tian, J.-H. Seo, D. V. Plant, Z. Ma, S. Fan, and W. Zhou, {\it Double-layer Fano resonance photonic crystal filters}, Opt. Express {\bf 21}, 24582 (2013).
\bibitem{Zhou2009} Y. Zhou, V. Karagodsky, B. Pesala, F. G. Sedgwick, and C. J. Chang-Hasnain, {\it A novel ultra-low loss hollow-core waveguide using subwavelength high-contrast gratings}, Opt. Express {\bf 17}, 1508 (2009).
\bibitem{Brueckner2010} F. Br\"uckner, D. Friedrich, T. Clausnitzer, M. Britzger, O. Burmeister, K. Danzmann, E.-B. Kley, A. T\"unnermann, and R. Schnabel, {\it Realization of a Monolithic High-Reflectivity Cavity Mirror from a Single Silicon Crystal}, Phys. Rev. Lett. {\bf 104}, 163903 (2010).
\bibitem{Huang2007} M. C. Y. Huang,  Y. Zhou, and C. J. Chang-Hasnain, {\it A surface-emitting laser incorporating a high-index-contrast subwavelength grating}, Nature Photonics {\bf 1}, 119 (2007).
\bibitem{Boutami2007} S. Boutami, B. Benbakir, X. Letartre, J. L. Leclercq, P. Regreny, and P. Viktorovitch, {\it Ultimate vertical Fabry-Perot cavity based on single-layer photonic crystal mirrors}, Opt. Express {\bf 15}, 12443 82007).
\bibitem{Wagner2016} T. Wagner, M. Sudzius, A. Mischok, H. Fr\"{o}b, and K. Leo, {\it Cross-coupled composite-cavity organic microresonators}, Appl. Phys. Lett. {\bf 109}, 043302 (2016).
\bibitem{Chen2010} L. Chen, H. Yang, Z. Qiang, H. Pang, L. Sun, Z. Ma, R. Pate, A. Stiff-Roberts, S. Gao, J. Xu, G. J. Brown, and W. Zhou, {\it Colloidal quantum dot absorption enhancement in flexible Fano filters}, Applied Physics Letters {\bf 96}, 083111 (2010).
\bibitem{Zhou2008} Y. Zhou, M. Moewe, J. Kern, M. C. Y. Huang, and C. J. Chang-Hasnain, {\it Surface-normal emission of a high-Q resonator using a subwavelength high-contrast grating}, Opt. Express {\bf 16}, 17282 (2008).
\bibitem{Sauvan2005} C. Sauvan, P. Lalanne, and J. P. Hugonin, {\it Slow-wave effect and mode-profile matching in photonic crystal microcavities}, Phys. Rev. B {\bf 71}, 165118 (2005).
\bibitem{Karagodsky2011} V. Karagodsky, C. Chase, and C. J. Chang-Hasnain, {\it Matrix Fabry--Perot resonance mechanism in high-contrast gratings}, Opt. Lett. {\bf 36}, 1714 (2011).
\bibitem{Bitarafan2017} M. H. Bitarafan and R. G. DeCorby, {\it On-Chip High-Finesse Fabry-Perot Microcavities for Optical Sensing and Quantum Information}, Sensors {\bf 17}, 1748 (2017).
\bibitem{Fattal2010} D. Fattal, J. Li,  Z. Peng, M. Fiorentino, and R. G. Beausoleil, {\it Flat dielectric grating reflectors with focusing abilities}, Nature Photonics {\bf 4}, 466 (2010).
\bibitem{Klemm2013} A. B. Klemm, D. Stellinga, E. R. Martins, L. Lewis, G. Huyet, L. O'Faolain, and T. F. Krauss, {\it Experimental high numerical aperture focusing with high contrast gratings}, Opt. Lett. {\bf 38}, 3410 (2013).
\bibitem{Arbabi2015} A. Arbabi, Y. Horie, A. J. Ball, M. Bagheri, and A. Faraon, {\it Subwavelength-thick lenses with high numerical apertures and large efficiency based on high-contrast transmitarrays}, Nature Communications {\bf 6}, 7069 (2015). 
\bibitem{Guo2017} J. Guo, R. A. Norte, and S. Gr\"{o}blacher, {\it Integrated optical force sensors using focusing photonic crystal arrays}, Opt. Express {\bf 25}, 9196 (2017).
\bibitem{Kemiktarak2012} U. Kemiktarak, M. Metcalfe, M. Durand, and J. Lawall, {\it Mechanically compliant grating reflectors for optomechanics}, Applied Physics Letters {\bf 100}, 061124 (2012).
\bibitem{Bui2012} C. H. Bui, J. Zheng, S. W. Hoch, L. Y. T. Lee, J. G. E. Harris, and C. W. Wong, {\it High-reflectivity, high-Q micromechanical membranes via guided resonances for enhanced optomechanical coupling}, Applied Physics Letters {\bf 100}, 021110 (2012).
\bibitem{Norte2016} R. A. Norte, J. P. Moura, and S. Gr\"oblacher, {\it Mechanical Resonators for Quantum Optomechanics Experiments at Room Temperature}, Phys. Rev. Lett. {\bf 116}, 147202 (2016).
\bibitem{Reinhardt2016} C. Reinhardt, T. M\"uller, A. Bourassa, and J. C. Sankey, {\it Ultralow-Noise SiN Trampoline Resonators for Sensing and Optomechanics}, Phys. Rev. X {\bf 6}, 021001 (2016).
\bibitem{Bernard2016} S. Bernard, C. Reinhardt, V. Dumont, Y.-A. Peter, and J. C. Sankey, {\it Precision resonance tuning and design of SiN photonic crystal reflectors}, Opt. Lett. {\bf 41}, 5624 (2016).
\bibitem{Yang2013} X. Yang, C. Husko, C. W. Wong, M. Yu, and D.-L. Kwong, {\it Observation of femtojoule optical bistability involving Fano resonances in high-$Q∕ V_m$ silicon photonic crystal nanocavities}, Appl. Phys. Lett. {\bf 91}, 051113 (2013).
\bibitem{Hui2013}  P.-C. Hui, D. Woolf, E. Iwase, Y.-I. Sohn, D. Ramos, M. Khan, A. W. Rodriguez, S. G. Johnson, F. Capasso, and M. Loncar, {\it Optical bistability with a repulsive optical force in coupled silicon photonic crystal membranes}, Appl. Phys. Lett. {\bf 103}, 021102 (2013)
\bibitem{Chen2017}  X. Chen, C. Chardin, K. Makles, C. Ca\"{e}r, S. Chua, R. Braive, I. Robert-Philip, T. Briant, P.-F. Cohadon, A. Heidmann, T. Jacqmin, and S. Deleglise, {\it High-finesse Fabry-Perot cavities with bidimensional Si3N4 photonic-crystal slabs}, Light: Science \& Applications {\bf 6}, e16190 (2017).
\bibitem{Thompson2008} J. D. Thompson, B. M. Zwickl, A. M. Jayich, Florian Marquardt, S. M. Girvin, and J. G. E. Harris, {\it Strong dispersive coupling of a high-finesse cavity to a micromechanical membrane}, Nature {\bf 452}, 72 (2008).
\bibitem{Kemiktarak2012NJP} U. Kemiktarak,  M. Durand, M. Metcalfe, and J. Lawall, {\it Cavity optomechanics with sub-wavelength grating mirrors}, New J. Phys. {\bf 14}, 125010 (2012)
\bibitem{Xuereb2012} A. Xuereb, C. Genes, and A. Dantan, {\it Strong Coupling and Long-Range Collective Interactions in Optomechanical Arrays}, Phys. Rev. Lett. {\bf 109}, 223601 (2012).
\bibitem{Piergentili2018} P. Piergentili, L. Catalini, M. Bawaj, S. Zippilli, N. Malossi, R. Natali, D. Vitali, and G. D. Giuseppe, {\it Two-membrane cavity optomechanics}, New J. Phys. {\bf 20}, 083024 (2018).
\bibitem{Gartner2018} C. G\"{a}rtner, J. P. Moura, W. Haaxman, R. A. Norte, and S. Gr\"{o}blacher, {\it Integrated optomechanical arrays of two high reflectivity SiN membranes}, arxiv:1809.06372 (2018).
\bibitem{Krause2012} A. G. Krause, M. Winger, T. D. Blasius, Q. Lin, and O. Painter, {\it A high-resolution microchip optomechanical accelerometer}, Nat. Photon. {\bf 6}, 768 (2012).
\bibitem{Cervantes2014} F. G. Cervantes, L. Kumanchik, J. Pratt, and J. M. Taylor, {\it High sensitivity optomechanical reference accelerometer over 10 kHz}, Appl. Phys. Lett. {\bf 104}, 221111 (2014).
\bibitem{Armata2017} F. Armata, L. Latmiral, A. D. K. Plato, and M. S. Kim, {\it Quantum limits to gravity estimation with optomechanics}, Phys. Rev. A {\bf 96}, 043824 (2017).
\bibitem{Qvarfort2017} S. Qvarfort, A. Serafini, P. F. Barker, and S. Bose, {\it Gravimetry through non-linear optomechanics}, Nature Comm. {\bf 9}, 3690 (2018).
\bibitem{Naesby2017} A. Naesby, S. Naserbakht, and A. Dantan, {\it Effects of pressure on suspended micromechanical membrane arrays}, Applied Physics Letters {\bf 111}, 201103 (2017).
\bibitem{Leinders2015} S. M. Leinders, W. J. Westerveld, J. Pozo, P. L. M. J. van Neer, B. Snyder, P. O'Brien, H. P. Urbach, N. de Jong, and M. D. Verweij, {\it A sensitive optical micro-machined ultrasound sensor (OMUS) based on a silicon photonic ring resonator on an acoustical membrane}, Sci. Rep. {\bf 5}, 14328 (2015).
\bibitem{Basiri2018} S. Basiri-Esfahani, A. Armin, S. Forstner, and W. P. Bowen, {\it Cavity optomechanical ultrasound sensing}, arxiv:1805.01940 (2018).
\bibitem{Islam2014} M. R. Islam, M. M. Ali, M.-H. Lai, K.-S. Lim, and H. Ahmad, {\it Chronology of Fabry-Perot Interferometer Fiber-Optic Sensors and Their Applications: A Review}, Sensors  {\bf 14}, 7451 (2014).
\bibitem{Fan2002} S. Fan and J. D. Joannopoulos, {\it Analysis of guided resonances in photonic crystal slabs}, Phys. Rev. B {\bf 65}, 235112 (2002).
\bibitem{Fan2003} S. Fan, W. Suh, and J. D. Joannopoulos, {\it Temporal coupled-mode theory for the Fano resonance in optical resonators}, J. Opt. Soc. Am. A {\bf 20}, 569 (2003).
\bibitem{Golub2006} I. Golub, {\it Distributed exponential enhancement of phase sensitivity and intensity by coupled resonant cavities}, Opt. Lett. {\bf 31}, 507 (2006).
\bibitem{Smith2003} D. D. Smith, H. Chang, and K. A. Fuller, {\it Whispering-gallery mode splitting in coupled microresonators}, J. Opt. Soc. Am. B {\bf 20}, 1967 (2003).
\bibitem{Moura2018} J. P. Moura, R. A. Norte, J. Guo, C. Sch\"{a}fermeie,r and S. Gr\"{o}blacher, {\it Centimeter-scale suspended photonic crystal mirrors}, Opt. Express {\bf 26}, 1895 (2018).
\bibitem{Nair2018} B. Nair, A. Naesby, B. R. Jeppesen, and A. Dantan, {\it Suspended silicon nitride thin films with enhanced and electrically tunable reflectivity}, arxiv:1809.08790 (2018).
\bibitem{Lee2002} J. Y. Lee, J. W. Hahn, and H.-W. Lee, {\it Spatiospectral transmission of a plane-mirror Fabry--Perot interferometer with nonuniform finite-size diffraction beam illuminations}, J. Opt. Soc. Am. A {\bf 19}, 973 (2002). 
\bibitem{Nair2017} B. Nair, A. Naesby, and A. Dantan, {\it Optomechanical characterization of silicon nitride membrane arrays}, Opt. Lett. {\bf 42}, 1341 (2017).
\bibitem{MIST} T. Germer, {\it Modeled Integrated Scatter Tool}, available at http://physics.nist.gov/scatmech.
\bibitem{Wu2018} S. Wu, J. Sheng, X. Zhang, Y. Wu, and H. Wu, {\it Parametric excitation of a SiN membrane via piezoelectricity}, AIP Advances {\bf 8}, 015209 (2018).
\bibitem{Midolo2018} L. Midolo, A. Schliesser, and A. Fiore, {\it Nano-opto-electro-mechanical systems}, Nature Nanotechnology {\bf 13}, 11 (2018).

\end{thebibliography}
\end{document}